\documentclass[twocolumn,superscriptaddress,preprintnumbers,amsmath,amssymb,showpacs,prl]{revtex4}

\newcommand{\IMSS}{Muon Science Laboratory and Condensed Matter Research Center, Institute of Materials Structure Science, High Energy Accelerator Research Organization, Tsukuba, Ibaraki 305-0801, Japan}
\newcommand{\Sokendai}{Department of Materials Structure Science, The Graduate University for Advanced Studies (Sokendai),  Tsukuba, Ibaraki 305-0801, Japan}
\newcommand{\MSLTokyo}{Materials and Structures Laboratory, Tokyo Institute of Technology, Yokohama, Kanagawa 226-8503, Japan}
\newcommand{\MCES}{Materials Research Center for Element Strategy, Tokyo Institute of Technology (MCES), Yokohama, Kanagawa 226-8503, Japan}

\newcommand{\msr}{$\mu$SR\xspace}
\newcommand{\Magazine}[4]{#1 {\bf #2}, #3 (#4).}
\newcommand{\JACS}{J. Am. Chem. Soc.\xspace}

\newcommand{\PRB}{Phys.~Rev. B\xspace}
\newcommand{\JPCC}{J.~Phys. Chem. C\xspace}
\newcommand{\JPSJ}{J.~Phys. Soc. Jpn.\xspace}

\newcommand{\lowElectron}{10$^{19}$cm$^{-3}$\xspace}
\newcommand{\highElectron}{10$^{21}$cm$^{-3}$\xspace}
\usepackage{txfonts}
\usepackage{graphicx} 
\usepackage{dcolumn} 
\usepackage{bm} 
\usepackage{xspace}
\usepackage{ulem} 
\usepackage[hypertex,colorlinks=true,linkcolor=black,citecolor=blue,filecolor=blue,urlcolor=blue,setpagesize=false,nesting=true]{hyperref}
\begin{document}

\title{Cage Electron-Hydroxyl Complex State as Electron Donor in Mayenite}

\author{M.~Hiraishi}
\affiliation{\IMSS}
\author{K.~M.~Kojima}
\affiliation{\IMSS}\affiliation{\Sokendai}
\author{M.~Miyazaki}
\affiliation{\IMSS}
\author{I.~Yamauchi}
\affiliation{\IMSS}
\author{H.~Okabe}
\affiliation{\IMSS}
\author{A.~Koda}
\affiliation{\IMSS}\affiliation{\Sokendai}
\author{R.~Kadono}\thanks{Corresponding author: ryosuke.kadono@kek.jp}
\affiliation{\IMSS}\affiliation{\Sokendai}
\author{S.~Matsuishi}
\affiliation{\MSLTokyo}
\author{H.~Hosono}
\affiliation{\MSLTokyo}\affiliation{\MCES}

\begin{abstract}
It is inferred from the chemical shift of muon spin rotation ($\mu$SR) spectra that muons implanted in pristine (fully oxidized) mayenite, [Ca$_{12}$Al$_{14}$O$_{32}]^{2+}$[$\square_5$O$^{2-}$] (C12A7, with $\square$ referring to the vacant cage), are bound to O$^{2-}$ at the cage center to form OMu$^-$ (where Mu represents muonium, a muonic analog of the H atom). However, an isolated negatively charged state (Mu$^-$, an analog of H$^-$) becomes dominant when the compound approaches the state of electride [Ca$_{12}$Al$_{14}$O$_{32}]^{2+}$[$\square_4$2e$^{-}$] as a result of the reduction process. Moreover, the OMu$^-$ state in the pristine specimen exhibits depolarization of paramagnetic origin at low temperatures (below $\sim$30~K), indicating that OMu$^-$ accompanies a loosely bound electron in the cage that can be thermally activated. This suggests that interstitial muons (and hence H) forming a ``cage electron-hydroxyl" complex can serve as electron donors in C12A7.
\end{abstract}

\pacs{72.80.Jc, 71.55.Ht, 76.75.+i}

\maketitle

Mayenite, Ca$_{12}$Al$_{14}$O$_{33}$ (C12A7), comprises a class of cage-structured compounds~\cite{Jeevaratnam_1964,Imlach_1971,Watauchi_JCrystGrowth2002} that has been drawing much attention since it was demonstrated to exhibit persistent photoconductivity upon hydrogenation~\cite{Hayashi_Nature2002} and to serve as an electride (in which the electron behaves as an anion)~\cite{Matsuishi_Science2003}. It is composed of calcium oxide (12CaO) and alumina (7Al$_2$O$_3$), which is made of the first, third, and fifth elements in the order of the Clarke number (natural abundance of the individual elements in the earth's crust), attracting broad interest in regard to ``green technology" where the replacement of rare chemical elements with useful functionalities with such common materials is the central focus.
\par
The chemical formula of a unit cell for mayenite in a fully oxidized state is represented by a positively charged lattice framework and two endohedral oxygen ions,  [Ca$_{24}$Al$_{28}$O$_{64}$]$^{4+}$[$\square_{10}$2O$^{2-}$] with $\square$ representing the vacant cage.  The unit cell has 12 cages, and two of them are occupied by O$^{2-}$ ions to maintain charge neutrality. The lattice structure belongs to a body centered cubic ($I\bar{4}3d$) with a lattice constant of 1.1989~nm~\cite{Palacios_InorgChem2008,Hayashi_NatCom2014}.
Because of the large free space ($\sim$0.4~nm) compared to the diameter of the O$^{2-}$ ion ($\sim$0.25~nm), the O$^{2-}$ ion is loosely bound to six Ca$^{2+}$ ions within the cage, virtually regarded as a free O$^{2-}$ ion.
Electric conductivity is achieved after hydrogenation, which leads to the substitution of an endohedral O$^{2-}$ ion with two H$^-$ ions, where an electron is dissociated by uv-light illumination, i.e.,
H$^-$ + $h\nu$ $\rightarrow$  H$^0$ + e$^-$~\cite{Hayashi_Nature2002}.
Interestingly, the O$^{2-}$ ions can be directly replaced with electrons (without anions) upon reduction treatment using Ca or Ti metals (C12A7:e$^-$)~\cite{Matsuishi_Science2003,Kim_NanoLett2007}. An electron introduced to the cage attracts two Ca$^{2+}$ ions on the $S_4$ symmetry axis of the cage to form a polaron state, and thus it is localized at the cage center~\cite{Sushko_PRL2003,Sushko_JACS2007,Kim_NanoLett2007}.
The polaron uses hopping motion to reach the neighboring cage by quantum tunneling, contributing to the conductivity.
As electron doping proceeds as described by the formula [Ca$_{24}$Al$_{28}$O$_{64}]^{4+}$[$2(1-x)$O$^{2-}+4x$(e$^-$)], it exhibits insulator-to-metal transition~\cite{Kim_NanoLett2007}, eventually exhibiting superconductivity with a transition temperature of $\sim$0.4~K in the compound with maximum doping ($x\simeq1$, corresponding to an electron concentration of $\sim$2$\times10^{21}$~cm$^{-3}$)~\cite{Miyakawa_JACS2007}.
\par
In spite of a relatively low work function ($\sim$2.4~eV, comparable to that of alkali metals), C12A7:e$^-$ is thermally and chemically stable in an ambient environment owing to the characteristic cage structure~\cite{Toda_AdvMater2007}, paving the way for a wide range of practical applications~\cite{Hayashi_Nature2002,Buchammagari_OrgLett2007,Ruszak_CatalLett2008,Kim_JACC2007,Adachi_MaterSciEngB2009,Toda_NatCommun2013,Kitano_NatChem2012,Kitano_NatCommun2015}. 
One of the most promising applications is highly efficient ammonia synthesis at a low energy achieved using Ru-loaded C12A7:e$^-$~\cite{Kitano_NatChem2012, Kitano_NatCommun2015}. This is expected to be an alternative method to the Haber-Bosch process, which contributes to more than 1~\% of the world's energy consumption.
However, there are a number of issues that still require clarification concerning the H kinetics in mayenite, e.g., whether or not the transient H$^+$ species exists as a result of the heterolytic H$_2$ dissociation on the Ru surface.  Considering the possibility that H would be present as an impurity (as is often the case in many semiconductors and metal oxides), microscopic information on the electronic structure (chemical state) of H in mayenite would be important for understanding a variety of H-related phenomena, including ammonia synthesis using Ru-loaded C12A7:e$^-$.
\par
Here, we report on the electronic structure of a muon as pseudo-hydrogen in mayenite with three different levels of electron doping, namely, pristine ($x=0$), \lowElectron, and \highElectron ($x\sim1$), investigated using muon spin rotation ($\mu$SR) spectroscopy.
$\mu$SR is a powerful technique for simulating the electronic structure of interstitial H because the implanted positive muon ($\mu^+$) can be regarded as a light radioisotope of a proton ($m_{\mu}\simeq 1/9m_{\rm p}$, with $m_p$ being the proton mass), simulating the chemical state of H by forming bound states with electrons. While their dynamical properties (e.g., diffusion) may differ considerably owing to the large mass difference, the local electronic structure of a muon is similar to that of H because of the negligible difference in the reduced electron mass ($\sim$0.5\%). Hereafter, we denote the chemical states of a muon as Mu$^+$, Mu$^0$, and Mu$^-$, which correspond to H$^+$, H$^0$, and H$^-$, respectively.
The present $\mu$SR results suggest that muons implanted in the pristine mayenite are bound to O$^{2-}$ ions at the cage center to form a complex state of a hydroxyl base and loosely bound electrons, ${\rm \{O}^{2-}\}+ {\rm Mu}\rightarrow {\rm \{OMu}^-$-e$^-\}$, whereas Mu$^-$ is formed in an electron-rich sample, $\{{\rm e}^-\}+ {\rm Mu}\rightarrow {\rm \{Mu}^-\}$ (where $\{\ \}$ indicates the cage).
Interestingly, the ${\rm \{OMu}^-$-e$^-\}$ complex state disappears when the temperature is increased beyond $\sim$30~K, indicating that the electron can be released by thermal activation, ${\rm \{OMu}^-$-e$^-\}\rightarrow {\rm \{OMu}^-\} + \{{\rm e}^-$\}. This suggests that an interstitial muon (and hence H) has a tendency of forming a ``cage electron-hydroxyl" complex state that can serve as an electron donor in mayenite.

\par
\begin{figure}[t]
 \centering
 \includegraphics[width=0.75\linewidth]{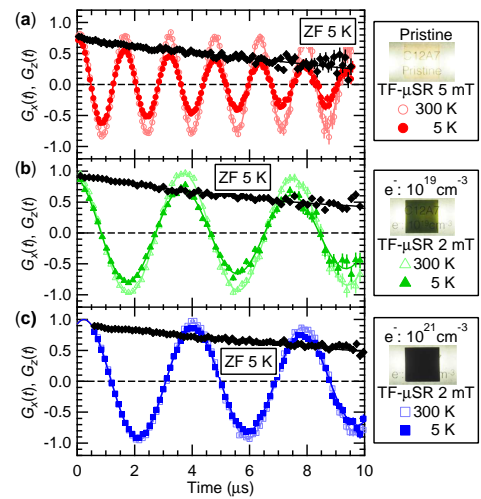}
 \caption{(Color online) Examples of wTF-\msr time spectra observed at 300~K and 5~K in mayenite with (a) pristine (5~mT), (b) \lowElectron (2~mT), and (c) \highElectron (2~mT) doping levels. ZF-\msr spectra observed at 5~K are also shown by filled diamonds to confirm the line shape of the envelop function. The solid curves are the best fits obtained using Eq.~(\ref{Eq_fit}). Pictures of respective samples are shown in the right column.}
 \label{Fig_spectra}
\end{figure}

Single crystalline samples of mayenite with three electron doping levels (0, \lowElectron, and \highElectron) were obtained using a reduction treatment. The details of sample preparation are described in an earlier report~\cite{Matsuishi_Science2003,Kim_NanoLett2007}.
As shown in Fig.~\ref{Fig_spectra}, the samples change color from transparent to green, and finally to black with increasing carrier concentration.
The electron doping levels were estimated in a semi-quantitative manner by comparing the colors of the samples with those of the previous specimens.
A conventional \msr experiment under a weak transverse field (wTF, $B_{\rm ext}\simeq$2--5 mT) was performed at the Paul Scherrer Institute (Switzerland) and at the J-PARC MUSE Facility.
For the chemical shift measurements, an additional \msr experiment on the pristine and \highElectron samples under a high transverse field (HTF, $B_{\rm ext}=6$~T) was performed at TRIUMF (Canada), where a muon beam ($\sim$100\% spin-polarized perpendicular to the beam momentum direction $\hat{z}$) was irradiated on single crystalline samples with a $(001)$ surface.
In these measurements, the applied external magnetic field was normal to the $(001)$ surface ($B_{\rm ext}\parallel \hat{y}$ for wTF and $\parallel \hat{z}$ for HTF).
The decay positron asymmetry for the transverse [$A_x(t)$] or zero/longitudinal (ZF/LF) field geometry [$A_z(t)$] was monitored by pairs of scintillation counters placed along the appropriate direction relative to the initial muon spin polarization.
\par
Figure \ref{Fig_spectra} shows typical examples of wTF-\msr time spectra that are normalized [$G_{x,z}(t)=A_{x,z}(t)/A_0$, with $A_0$ being the full instrumental asymmetry calibrated by silver specimen] in the respective samples.  All the spectra exhibit damping oscillations with a frequency expected for the diamagnetic muon state (i.e.,  Mu$^+$ or Mu$^-$), $\omega=\gamma_\mu B_{\rm ext}$, where $\gamma_\mu$ ($=2\pi\times 135.53$ MHz/T) is the muon gyromagnetic ratio.
While the spectra exhibit negligibly slow depolarization at 300~K, the depolarization rate is enhanced at 5~K, with a tendency of greater depolarization for a lower carrier concentration.
As is confirmed in the ZF-\msr spectra at 5 K (filled diamonds in Fig.~\ref{Fig_spectra}), the envelop function exhibits an exponential behavior (particularly in the earlier time period), suggesting that depolarization is induced by an unpaired electron(s).
It is also noticeable (see below) that the pristine and \lowElectron samples exhibit a slight reduction of the initial polarization at low temperatures [i.e., $G_{x,z}(0)<1$].
Considering these factors, we used the following form for the curve fit of the wTF-$\mu$SR spectra:
\begin{align}
  A_x(t)&=\left[A_{\rm s}(T)e^{-\delta_{\rm n}^2t^2}\{(1-p)+pe^{-\lambda t}\}+A_{\rm BG}\right]\cos(\omega t+\phi),
  \label{Eq_fit}
\end{align}
where $\delta_{\rm n}$ and $\lambda$ are the depolarization rate due to nuclear magnetic moments and unpaired electrons, respectively, $\phi$ is the initial phase of precession, $A_s(T)$ and $A_{\rm BG}$ are the initial asymmetry for signals from muons stopped in the sample (with $T$ being the measured temperature) and a $T$-independent background from the sample holder, respectively, and $p$ is the fraction exhibiting exponential damping.
Because of the low depolarization rate, $p$, $\delta_{\rm n}$ and $\lambda$ were best determined at the lowest temperature for each sample, and the values of $p$ and $\delta_{\rm n}$ were fixed (see Table.~\ref{tbl:f_delta}) for the analysis in order to detremine $\lambda$ reliably over the entire $T$ range.
\par

\begin{table}[!t]
	\caption{Values of parameters in Eq.~(\ref{Eq_fit}) that were presumed to be independent of temperature, which were deduced from curve fits of the spectra obtained at 5 K.}
	\centering
	\vspace{1mm}
  \begin{tabular*}{70mm}{@{\extracolsep{\fill}}cccc}
		\hline\hline
		& Pristine &  e$^-$:10$^{19}$cm$^{-3}$ &  e$^-$:10$^{21}$cm$^{-3}$\\
    \hline
    $p$ & 0.360(7) & 0.49(7) & 0.49(3)\\
    $\delta_{\rm n}$ [$\mu$s$^{-1}$] & 0.080(2) & 0.055(8) & 0.044(3)\\
    \hline\hline
 	\end{tabular*}
	\label{tbl:f_delta}
\end{table}

\begin{figure}[t]
 \centering
  \includegraphics[width=0.65\linewidth,clip]{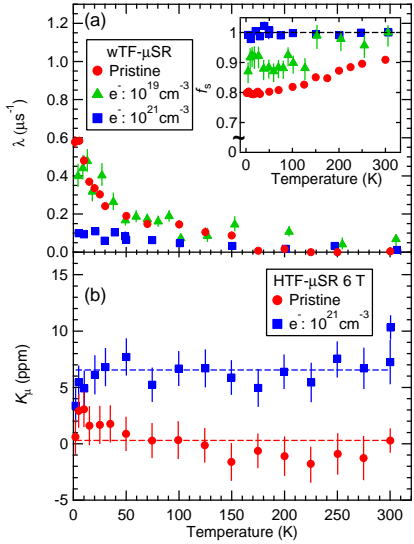}
  \caption{(Color online) Temperature dependence of (a) muon depolarization rate deduced from wTF-\msr measurements and (b) frequency shift ($K_\mu$) deduced from HTF-\msr measurements in mayenite with different levels of electron doping.  The dashed lines in (b) are the best fit assuming a temperature-independent $K_{\mu}$. Inset: fractional yield of \msr signal at $t=0$ with respect to the value corresponding to 100\% Mu$^{+/-}$ formation (see text).}
  \label{Fig_rlx_asy_shift}
\end{figure}

The temperature dependence of $\lambda$ is shown in Fig.~\ref{Fig_rlx_asy_shift}(a), where $\lambda$ in the pristine sample exhibits a gradual increase as the temperature decreases, and its slope increases below $\sim$30~K.
A similar behavior is observed in the \lowElectron sample with a diminished slope, but is absent in the \highElectron sample.
Apart from this, as shown in the inset of Fig.~\ref{Fig_rlx_asy_shift}(a), the behavior of the signal fraction, $f_{\rm s}\equiv A_{\rm s}(T)/(A_0-A_{\rm BG})$, indicates that a few of the implanted muons ($\le$20\%) are instantaneously depolarized (i.e., $\lambda\gg10^1$ MHz) at lower temperatures in the pristine and \lowElectron samples. Such a fast depolarization may be attributed to the fractional formation of the Mu$^0$ state  (see the discussion below).

\par
More detailed information on the chemical state of muon can be obtained by high-precision frequency shift measurements. 
The time spectra under $B_{\rm ext}=6$~T were analyzed by curve fits using Eq.~(\ref{Eq_fit}) (with $A_{\rm BG}\simeq0$ for this experimental condition) to yield the shift
\[K_{\mu}=\frac{\omega-\gamma_{\mu}B_{\rm ext}}{\gamma_{\mu}B_{\rm ext}},\]
where $B_{\rm ext}$ was simultaneously monitored by the frequency of muons stopped in the sample holder  (made of CaCO$_3$ which is a non-magnetic insulator and serving as a muon counter) just behind the sample.

Fig.~\ref{Fig_rlx_asy_shift}(b) shows the temperature dependence of $K_{\mu}$ observed in the pristine and \highElectron samples.  It is mostly independent of temperature for both the samples, and the temperature-averaged $K_{\mu}$ is +0.3(4)~ppm and +6.6(4)~ppm in the respective samples.
These values coincide with the proton chemical shift in relevant compounds observed by $^1$H-NMR, namely -0.8~ppm in C12A7:OH$^-$ and +5.1~ppm in C12A7:H$^-$ \cite{Hayashi_NatCom2014,Yoshizumi_JPCC2012}. Thus, we can assign the chemical state of diamagnetic muon in the pristine and \highElectron samples to OMu$^-$ and Mu$^-$, respectively.
However, assuming that the contribution of diamagnetic shielding to hyperfine coupling is common between muon and proton with its magnitude enhanced by $\gamma_\mu/\gamma_{\rm p}=3.1832$ (where $\gamma_{\rm p}=2\pi\times42.5774$ MHz/T is the proton gyromagnetic ratio), $K_{\mu}$ in the electron-doped sample is considerably smaller than that expected from $^1$H-NMR ($5.1\times3.18\simeq16.2$~ppm).  The discrepancy is tentatively attributed to the metallic environment of Mu$^-$ that is largely different for H$^-$ in fully hydrogenated (insulating) C12A7:H$^-$.



\par
\begin{figure}[t]
 \centering
 \includegraphics[width=0.85\linewidth, clip]{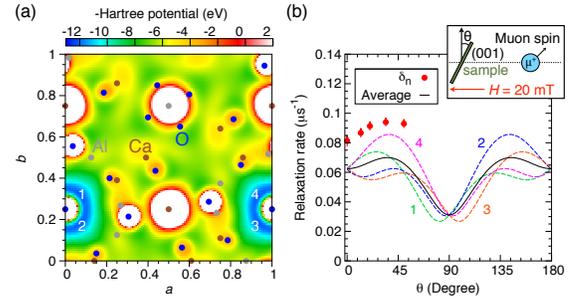}
 \caption{(Color online) (a) Hartree potential map at $z=0.378$ plane in pristine mayenite, where a free O atom is placed at the center of a cage $(0,0.25,0.375)$. The brown, gray, and blue dots represent the Ca, Al, and O atoms, respectively, within $0.25\le z\le0.5$. Potential minima for muon (shown by blue hatched area) are circularly spread around O. (b) Angle dependence of muon spin relaxation rate ($\delta_{\rm n}$) under a transverse field of 20~mT. Dashed curves are simulated $\delta_{\rm n}$ for muons located at the position of 1$\sim$4, with their average represented by a solid curve. Inset: Schematic view of the relation between the magnetic field and sample rotation angle $\theta$ from (001) surface.}
  \label{Fig_muonsite}
\end{figure}

Since the wTF-\msr spectra in the pristine sample exhibit sizable $\delta_{\rm n}$ (primarily due to the nearest neighboring $^{27}$Al nuclei), the local environment of muon can be further examined by comparing $\delta_{\rm n}$ with that estimated for the presumed muon sites.  For narrowing down the candidate muon sites, the first principle calculation using the Vienna Ab-initio Simulation Package (VASP)~\cite{VASP} was performed for pristine mayenite.
Fig.~\ref{Fig_muonsite}(a) shows a contour map of the Hartree potential at $z=0.378$ calculated with a free oxygen placed at the center of a cage $(0, 0.25, 0.375)$ with other cages left open ($120\times120\times120$ meshes),
where the potential minima are circularly spread around the free O$^{2-}$ at the cage center.
The potential contour exhibits monotonous increase in departing from $z=0.378$ along the $c$-axis due to the presence of Al$^{3+}$ ions, and thus the potential minima comprise a toroid in the three-dimensional space.
The distance between the oxygen and toroid is $\sim$0.12~nm, which corresponds to the typical OMu (OH) bond length in other oxides.

\par
It is known that $\delta_{\rm n}$ depends on the direction of $B_{\rm ext}$ with respect to the crystalline axis [defined by the polar angle $\theta$, see the inset of Fig.~\ref{Fig_muonsite}(b)], and thus the $\theta$ dependence of $\delta_{\rm n}$ is useful in locating muon sites.
In the analysis for the angle-resolved data,  $\delta_{\rm n}$ was extracted by the curve-fits using Eq.~(\ref{Eq_fit}) with $\lambda$ and $p$ assumed to be independent of $\theta$.
The obtained  $\delta_{\rm n}$ vs. $\theta$ are shown in Fig.~\ref{Fig_muonsite}(b), where the dashed curves represent the simulation results with $\delta_{\rm n}^i$ $(i=1,2,3,4)$ corresponding to the $i$-th muon site indicated in Fig.~\ref{Fig_muonsite}(a), and the solid line represents their average $\overline{\delta}_{\rm n}=[\sum_{i=1}^4(\delta_{\rm n}^i)^2/4]^{1/2}$.
While $\delta_{\rm n}$ exhibits systematically larger values than $\overline{\delta}_{\rm n}$ (probably due to uncertainty originating from imperfect separation of two components in the curve-fit), its angle dependence is qualitatively in line with the simulation, providing additional support that muons implanted into the pristine sample are bound to oxygen at the cage center.


Now, let us discuss the electronic structure of muon in mayenite.
The temperature independence of $\lambda$ and $f_{\rm s}$ in the \highElectron sample implies that Mu$^-$ is stable in the entire temperature range of the present study (5-300~K), consistent with the earlier suggestion that H$^-$ is more stable than H$^0$ (with e$^-$ in other cages) \cite{Yoshizumi_JPCC2012,Hayashi_BullChemSocJ2007}.
In contrast, the pristine compound has two free O$^{2-}$ ions in a unit cell, where muons (positively charged during the thermalization process immediately after implantation) tend to be attracted by the negative charge of O$^{2-}$ to form a muonic analog of hydroxyl base, OMu$^-$, with one extra electron left unpaired in the equilibrium state [i.e., $\{{\rm O}^{2-}\}$ + Mu$^+$  + e$^-$ $\rightarrow$  $\{{\rm OMu}^-\}$ + $\{{\rm e}^-\}$, where the electrons necessary for charge compensation are provided to muons (up to a fraction $p$) from short-lived epithermal electrons generated near the end of the muon radiation track].
It is inferred from the Bader charge analysis using VASP~\cite{Bader} that introduction of oxygen into the cage center leads to a valence of O$^{-1.23}$, suggesting that the monovalent state (OH$^-$) is more stable than the divalent state (O$^{2-}$), which also supports the formation of the OMu$^-$ base.
We also analyze the Bader change of hydrogen introduced into the cage center, resulting in negative charge of H$^{-0.73}$. This also supports the existence of a stable Mu$^-$ in electron-doped samples.

\begin{figure}
 \centering
 \includegraphics[width=0.8\linewidth]{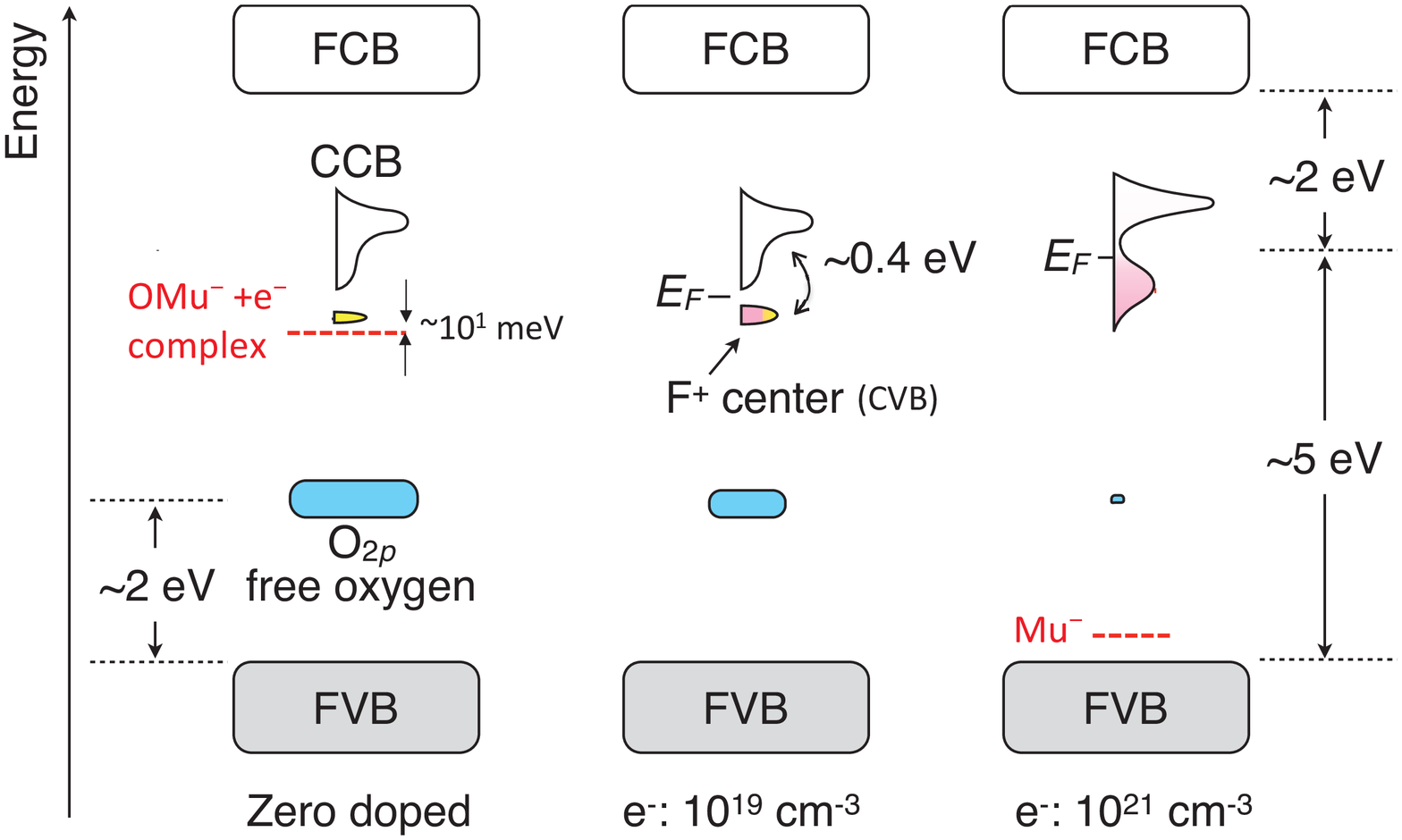}
 \caption{(Color online) Schematic energy diagrams of mayenite \cite{Matsuishi_JPCC2008,Souma_JPSJ2010,McLeod_PRB2012}. The blue (pink) and yellow hatched regions show the occupied levels of oxygen (electron) in the cage center. The red dashed lines show the electronic levels associated with muon. FCB (FVB) and CCB (CVB) denote the frame conduction (valence) band and cage conduction (valence) band, respectively. }
  \label{Fig_band}
\end{figure}

The increase of $\lambda$ at low temperatures in the pristine and \lowElectron compounds strongly suggests that the unpaired electron is localized near the OMu$^-$ base to form a complex state in a cage, $\{{\rm OMu}^-$-e$^-\}$. This seems to be in marked similarity with the situation recently observed for muons in rutile TiO$_2$ at low temperatures, where the unpaired electron is loosely bound to the Ti$^{3+}$ ion next to the OMu$^-$ base to form a Ti-O-Mu complex state \cite{Shimomura:2015}.
In this regard, the decrease of $f_{\rm s}$ with decreasing temperature in the pristine and \lowElectron compounds may provide additional support for the presence of electrons near OMu$^-$ since the fast depolarization can be attributed to the formation of Mu$^0$ by the reverse process, $\{{\rm OMu}^-$-e$^-\}$ $\rightarrow$ $\{{\rm Mu}^0\}$ + $\{{\rm O}^{2-}\}$.
The absence of Mu$^0$-like precession signal suggests that the magnitude of the hyperfine parameter for Mu$^0$ isolated in the cage may be comparable to a vacuum value of 4.463~GHz, which exceeds the present time resolution, resulting in the missing fraction.
\par
The energetics of Mu/H in the band structure of mayenite may be understood in the schematic diagrams shown in Fig.~\ref{Fig_band}. According to the earlier reports~\cite{Kamiya_JAPP2005,Sushko_JACS2007,Toda_AdvMater2007,Matsuishi_JPCC2008,Souma_JPSJ2010,McLeod_PRB2012}, the conduction band consists of the frame conduction band (FCB) and cage conduction band (CCB), where CCB is associated with electrons at the cage center.
In pristine mayenite, CCB is unoccupied while free O$^{2-}$-2$p$ levels form the occupied state located at 2~eV above the frame valence band (FVB). Slight electron doping divides the CCB to an empty band and fully occupied cage valence band (CVB, similar to the F$^+$ center observed in CaO~\cite{Henderson:68}) due to the polaron formation~\cite{Kamiya_JAPP2005,Matsuishi_JPCC2008,Souma_JPSJ2010}. By further electron doping, the kinetic energy gain surpasses the lattice distortion to induce the insulator-to-metal transition~\cite{Kim_NanoLett2007}.
Therefore, it is presumed that the OMu$^-$ hydroxyl base in pristine mayenite serves as an electron donor to the vacant cage to form the F$^+$-like center (i.e., $\{{\rm OMu}^-$-e$^-\}\rightarrow\{{\rm OMu}^-\} + \{{\rm e}^-\}$).  The activation energy for the electron release is estimated to be 2.7(2)~meV by the curve fit of $\lambda$ shown in Fig.~\ref{Fig_rlx_asy_shift}(a) using the data below 100~K~\cite{Cox_JPhysCondMat2006}. This provides the microscopic origin of photo-induced conductivity in mayenite after H$^+$ ion implantation~\cite{Miyakawa:03}, where $\{{\rm H}^-\}$ would be generated via the process, $\{{\rm O}^{2-}\}$ + 2H$^0$ $\rightarrow$  $\{{\rm OH}^-\}$ + $\{{\rm e}^-\}$ +H$^0$ $\rightarrow$  $\{{\rm OH}^-\}$ + $\{{\rm H}^-\}$. 

Meanwhile, the electronic level associated with Mu$^-$ in the \highElectron sample would be situated deep in the band gap, as is suggested from the photon energy ($h\nu\simeq4$ eV) required for inducing the reaction $\{{\rm H}^-\}$ + $h\nu$ $\rightarrow$  $\{{\rm H}^0\}$ + $\{{\rm e}^-\}$~\cite{Hayashi_Nature2002}. Considering that there still remain many vacant cages available for muon even in \highElectron sample, the present result indicates that muon (as well as H) is unstable as an isolated Mu$^+$ or Mu$^0$ in mayenite.

\par
In conclusion, this study has shown that muon in mayenite forms either a hydroxyl-like OMu$^-$ state or an isolated Mu$^-$ state, depending on the degree of reduction.
The enhancement of muon spin depolarization rate below $\sim$30~K in the pristine sample suggests the presence of a cage electron-OMu$^-$ complex state, where the activation energy for the bound electron as low as 10$^1$~meV suggests that the complex state serves as an electron donor.
The result in fully electron-doped samples indicates that Mu$^-$ (and hence H$^-$) state is stable, and transient H$^+$ that breaks the balance of hydrogen storage-release reaction in Ru-loaded C12A7:e$^-$ is unlikely to exist.
\par
We would like to thank the staff of J-PARC MUSE, PSI, and TRIUMF for their technical support during the $\mu$SR experiment. This work was supported by MEXT Elements Strategy Initiative to Form Core Research Center. HH thanks a financial support from the ACCEL program of the Japan Science and Technology Agency (JST). The muon experiment was partly supported by the Inter-University Research Program of IMSS, KEK (Proposal No.~2012B0221 and 2013A0111).
The muon site calculation and the Bader charge analysis were performed under the KEK Large Scale Simulation Program No.14/15-13(FY2014-2015).
\par

\end{document}